\documentclass[final]{elsarticle}

\usepackage{amsmath}
\usepackage{amssymb}
\usepackage{pstricks}
\usepackage{natbib}
\usepackage{psfrag}
\newcommand{\vel}{\mathbf{u}}
\newcommand{\uv}{\vel(t,\mathbf{x})}
\newcommand{\duv}{\dot{\vel}(t,\mathbf{x})}
\newcommand{\ev}{\mathbf{e}_n(t,\mathbf{x})}
\newcommand{\vv}{\mathbf{v}}
\newcommand{\evni}{\mathbf{e}_n(t,\mathbf{x_i})}
\newcommand{\uvi}{\vel(t,\mathbf{x_i})}
\newcommand{\opl}{\mathcal{D}_{H\!a}}
\newcommand{\ha}{H\!a}
\newcommand{\opf}{OpenFOAM}

\journal{Journal of Computational Physics}

\begin{document}
\begin{frontmatter}


\title{Spectral Direct Numerical Simulations of low Rm MHD channel flows based on the least dissipative modes}

\author{K. Kornet and A. Poth\'erat}
\address{Applied Mathematics Research Centre, Coventry University, Priory street, Coventry CV1 5FB, UK}
\begin{abstract}
We put forward a new type of spectral method for the direct numerical
simulation of flows where anisotropy or very fine boundary layers are present.
The mean idea is to take advantage of the fact that such structures are
dissipative and that their presence should reduce the number of degrees of
freedom of the flow, when paradoxically, their fine resolution incurs extra
computational cost in most current methods. The principle of this method is to
use a functional basis with elements that already include these fine structure
so as to avoid these extra costs.  This leads us to develop an algorithm to
implement a spectral method for arbitrary functional bases, and in particular,
non-orthogonal ones.  We construct a basic implementation of this algorithm to
simulate Magnetohydrodynamic (MHD) channel flows with an externally imposed,
transverse magnetic field, where very thin boundary layers are known to develop
along the channel walls. In this case, the sought functional basis can be built
out of the eigenfunctions of the dissipation operator, which incorporate these
boundary layers, and it turns out to be non-orthogonal. We validate this
new scheme against numerical simulations of freely decaying MHD turbulence
based on a Finite Volume code and it is found to provide accurate results. Its
ability to fully resolve wall-bounded turbulence with a number of modes close
to that required by the dynamics is demonstrated on a simple example. This
opens the way to full blown simulations of MHD turbulence under very high
magnetic fields, which were until now too computationally expensive, as the computational
cost of the proposed method, in contrast to traditional methods,  does not depend on the intensity of the magnetic field.

\end{abstract}

\end{frontmatter}
\section{Introduction}
In the field of numerical simulations of fluid flows, Spectral Methods present
well established advantages in terms of precision and resolution over classical
mesh-based methods such as finite differences, finite elements or finite
volumes methods. On the flip side, they are only well developed and widely used
for a limited set of functional bases (such as Fourier bases), and therefore
tend to be limited to ideal geometries. 
Difficulties start appearing as soon as bounding walls are present: in a
channel bounded by two parallel walls, Tchebychev polynomials are used in the
transverse direction that present the same advantages as Fourier modes in
periodic domains (orthogonality in both continuous and discrete physical
domains, fast forward and inverse transforms between real and spectral spaces
available to calculate non-linear terms, and exponential convergence). They are
particularly well suited to satisfy no-slip boundary conditions at the wall
because the locus of their zeros, used as collocation points, are all the more
concentrated near the walls as their order is high. This property, most welcome
to resolve strong velocity gradients of wall boundary layers, is also their
downfall.  When the boundary layers become very thin, the number of basis
elements required to resolve them becomes very large and the spatial
distribution of the collocation points becomes highly inhomogeneous. This
incurs high computational costs.\\ One such example
is found in flows of liquid metals under high magnetic fields (such as in the
cooling blankets of ITER, the future nuclear fusion reactor
\cite{smolentsev10_fed}). The corresponding idealised configuration is that  of
a liquid metal channel flow (height $h\sim10\,\mathrm{cm}$) placed in a
homogeneous static transverse magnetic field $B\sim10\,\mathrm{T}$: Hartmann
wall boundary layers are known to develop and their thickness scales as
$\delta\sim h\sqrt{\rho\nu/\sigma}/ B$. For a typical liquid metal of
conductivity $\sigma\simeq 10^6\,\mathrm{S/m}$, density
$\rho\sim10^4\,\mathrm{kg/m^3}$ and kinematic viscosity
$\nu\sim10^{-7}\,\mathrm{m^2/s}$, $\delta \sim 10^{-7}-10^{-6}\,\mathrm{m}$.
The number of Tchebychev polynomials required to resolve such a flow scales
with the Hartmann number $\ha \sim h/\delta$, and demands unrealistic
computational power \cite{boeck07}. Yet, the enormous dissipation incurred by
friction and Joule dissipation in these layers decimates the degree of freedom
of the flow when $\ha$ becomes large. Their number can be estimated through the
dimension $d_M$ of the attractor of the underlying system, for which an upper
bound was shown to scale as $\ha^{-1}$ \cite{pa06_pf}. The fact that the number
of modes needed to resolve the flow completely increases in numerical
simulations at high $\ha$ is therefore a property of the spectral method based
on these polynomials, but does not reflect any physical constraint.\\
An "ideal" spectral method would only require a number of modes of the order of the attractor dimension, which would be achieved by using a functional basis spanning the smallest linear space that embeds the attractor. Finding such basis in 
complex realistic case is unfortunately not a realistic task. \cite{pa06_pf} 
however proved that in the case of the MHD channel flow, the basis of 
eigenmodes of the linear part of Navier-Stokes equations spanned a space in 
which any $d_M-$ dimensional volume of initial conditions contracted to 0 
through the governing equations. From the physical point of view, these modes 
were shown to exhibit numerous properties of the actual MHD channel flow:
 same anisotropy and, especially, their profile across the channel includes that of a Hartmann layer, so representing these layers using this basis only 
requires a small number of modes.\\
This paper explores the idea of building a numerical spectral code based on
this "dynamically relevant" functional basis. Expansions over eigenbases of
linear operators have been commonly used to solve viscous flows
\cite{ladyzhenskaya68} and other problems involving the Laplacian operator.
Some problems also require tailored functional bases to deal with specific
geometries and boundary conditions: in the context of geophysical flows,
\cite{livermore07_jcp} put forward  
 a basis to circumvent difficulties arising from the singularities of 
spherical harmonics at the center of a spherical domain. The novelty here is to 
apply this technique to highly non-linear problems in the hope of keeping the 
number of modes required for full resolution close to that strictly required by 
the dynamics of the system. Since current spectral codes rely on a number of 
favourable properties of classical Fourier or Tchebychev bases (orthogonality, 
fast transforms), building a new code for arbitrary functional bases poses a
number of problems, particularly in dealing with the non-linear part of the
Navier-Stokes equations. The absence of these features makes the implementation
more difficult and adversely affects performance. Nevertheless, the challenge
of using dynamically relevant bases is to outweigh this loss by using 
significantly less modes. For MHD channel flows, this strategy is expected to 
become advantageous at high $\ha$. Therefore, the purpose of the present paper 
is to demonstrate an implementation of this method for MHD channel flows, to 
show that the resulting code provides accurate results and also that it makes 
it possible to simulate regimes of very high magnetic fields otherwise not 
accessible. The emphasis is laid on the generality of this implementation, 
rather than performance optimisation. The code we shall describe can be 
used with a large variety of other functional bases associated to different 
geometries and boundary conditions.\\
The governing equations for the MHD channel flow and the associated functional
base are presented in Section \ref{sec:equations}. The numerical method is
described in  section \ref{sec:method}. Finally, the code is tested in section
\ref{sec:tests}, where its ability to resolve freely decaying turbulence in an
MHD channel is compared to that of a finite volume code. We also show on an
example that flows at high Hartmann numbers can be resolved with limited
computational power.\\
\section{Governing equations and their properties}
\label{sec:equations}
\subsection{The Low-Rm MHD equations}
\begin{figure}
\centering
\psfrag{U}{$\mathbf U$}
\psfrag{B}{$\mathbf B$}
\psfrag{ex}{$\mathbf e_x$}
\psfrag{ey}{$\mathbf e_y$}
\psfrag{ez}{$\mathbf e_z$}
\psfrag{al}{$\alpha$}
\psfrag{L}{$L$}
\includegraphics[width=8cm]{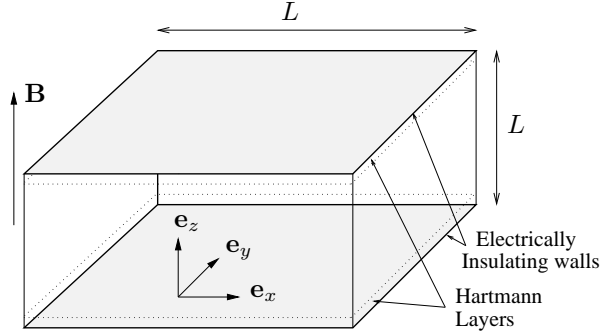}
\caption{Geometry of the Channel flow with transverse magnetic field}
\label{fig:channel}
\end{figure}

Flows of liquid metals in engineering applications are usually described within the frame of the Low Magnetic Reynolds number ($R\!m$) approximation. This 
applies to problems where the flow is neither intense nor conductive enough to 
induce a magnetic field comparable to an externally applied one. The full 
system of the induction equation and the Navier-Stokes equations for an 
incompressible fluid are then approximated to the first order in $R\!m$, which 
represents the ratio of these two fields. This leads to the following system \cite{roberts67}:
%
\begin{eqnarray}
\frac{\partial \vel}{\partial t} + (\vel \cdot \nabla) \vel  &=& 
- \frac{1}{\rho} \nabla p + \nu \Delta\vel  + \mathbf{j} \times \vel \,, \label{eq:nav}\\
\nabla \cdot \vel &=& 0 \,, \label{eq:cont}\\
\nabla \cdot \mathbf{j} &=& 0 \,, \label{eq:contj}\\
\mathbf{j} &=& \sigma ( - \nabla \Phi + \vel \times \mathbf{B}) \,, \label{eq:ohm}
\end{eqnarray}
where $\vel$ denotes fluid velocity, $\mathbf{B}$ - magnetic field,
$\mathbf{j}$ - electric current density, $\nu$ - kinematic viscosity, $\sigma$ -
electrical conductivity, $\Phi$ - electric potential.  We consider a channel
flow with a homogeneous transverse magnetic field $B\mathbf e_z$ and impermeable ($\vel|_{wall} =
\mathbf{0}$), electrically insulating ($\mathbf{j \cdot n}|_{wall} =
\mathbf{0}$) walls located at $z=\pm L$.  In the $xy$ directions we adopt the
periodic boundary conditions with period $2 \pi L$ (see figure \ref{fig:channel}). Under this assumptions and
using  the reference scale $L$, time $L^2/\nu$ and velocity $\nu/L$ the above
set of equations can be expressed in dimensionless form:
\begin{equation}
\frac{\partial \vel}{\partial t} + P(\vel \cdot \nabla) \vel = 
\Delta\vel - \frac{1}{\ha^2} \Delta^{-1}\partial_{zz} \vel \,,
\label{invlapl}
\end{equation}
where $\ha=LB\sqrt{\sigma/\rho\nu}$ is the Hartman number and $P$ denotes
orthogonal projection onto the subspace of solenoidal fields. $\Delta^{-1}\partial_{zz} \vel$ is defined as a vector field $\mathbf{w}$
which is a solution to equation:
\[
\Delta \mathbf{w} = \partial_{zz} \vel \,.
\]
The boundary conditions necessary to solve the above equation are derived from the boundary condition for the electric current.
Namely, we require that $w$ be periodic in $xy$ directions and that at walls $z=\pm1$,
\[
\nabla \times \mathbf{w} = \nabla \times (\vel \times \mathbf{e}_z) \,.
\]

Even though $\nu/L$ provides a natural scale for velocity, actual velocities,
particularly in turbulence flows are most of the time significantly higher. If
$U_0$ is their typical scale, the Reynolds number $U_0 \nu/L$ provides a
measure of the intensity of turbulence.

\subsection{Eigenvalue problem for the dissipation operator}
\label{eigmodes}

In \cite{dyp09_tcfd}, a set of solenoidal solutions to eigenvalue problem
of the operator $\mathcal L$ that represents the linear part of eq. (\ref{invlapl}):

\begin{eqnarray}
    \opl \mathbf u= \frac{1}{\ha^2} \Delta^{-1}\partial_{zz} \vel - \Delta\vel &=& \lambda \vel \label{eigen}\,,\\
    \nabla \cdot \vel &=& 0 \label{divfree}
\end{eqnarray}
together with the problem's boundary conditions, was derived (for their full
form see \ref{appeq}). The features of flows at high $\ha$ are strongly
determined by the properties of this operator. Because of this, the set of
modes built out of its eigenfunctions elements includes structures that
are actually present in the flow. Laminar and turbulent Hartmann boundary
layers that develop along the channel walls appear, in particular, as built-in
features of these modes \cite{dyp09_tcfd,pdy10_jfm}. They are therefore natural
candidates to be used as elements of a functional basis in a numerical spectral
scheme.  Moreover, these modes all have negative eigenvalues, and it can be
shown that to resolve the flow completely, it is only necessary to take into
account all modes with eigenvalue $\lambda$ with a modulus below a maximum
$|\lambda_{\rm max}|$, such that their total number scales as $Re^2/\ha$
\cite{pa06_pf}. Since the operator defined in (\ref{eigen}) represents the sum
of viscous and Joule dissipation, the set of modes defined in this way is in
fact the set of \emph{least dissipative modes}. For sufficiently large values
of $\ha$, this number becomes significantly smaller than the number of Fourier
or Tchebychev  modes necessary to resolve the Hartmann layers
\citep{dyp09_tcfd}.  However there are also some disadvantages in using these
alternative modes. The main complication comes in calculation of nonlinear
terms. In standard spectral schemes using an orthogonal basis, these are first
"transformed back" to be expressed in terms of physical coordinates $(x,y,z)$.
The non-linear terms are then calculated in this expansion and subsequently
their spectral expansion is found by calculating their scalar products with
each of the basis elements. The last step uses the orthogonality of the modes.
The eigenbasis of $\opl$, by contrast, is not orthogonal. In
theory, projection can still be performed without an orthogonal basis, using
the basis of eigenfunctions and associate eigenfunctions of the adjoint
operator. The latter may, however, be difficult to express explicitly, as in
the present case. Therefore we chose to develop an alternative procedure for 
projection of non linear terms on the space spanned by these modes, which we 
shall now describe.
A significant advantage of using a decomposition over the eigenmodes of
$\opl$ is that since these modes satisfy (\ref{eq:cont}) and
(\ref{eq:contj}), the solution conserves mass and charge to the precision of
the truncation error by construction (throughout the paper, we will refer to
this level of precision as "exact" in the sense that it involves no error
originating from the algorithm). Conservation of charge is particularly crucial
in problems with electrically insulating boundaries, where eddy currents must
close withing the fluid domain. Failure to ensure it to the precision of the
algorithm results in numerical instability and/or significant errors in the
final solution, as proved by \cite{ni07_jcp} for codes based on the finite
volume method. In such codes, dedicated algorithms have to be designed  to take
care of this particular difficulty that can incur extra computational cost (see
\cite{dp12_jfm}).

\section{Method}
\label{sec:method}
Given a complete set of basis functions spanning a space of solutions to the
considered equations, any velocity field can be expressed in two ways. In the
first one, which we call a physical expansion, it is expressed as a function of
coordinates $(x,y,z)$. In the second one, which we call a spectral expansion, 
it is
expressed as a series over the set of basis functions.  The idea behind
spectral methods is to search for a solution of the governing equations in its
spectral expansion.  Our aim  is to use a basis consisting of functions that
are better suited for the considered physical flow then the usually adopted
Fourier or Tchebychev expansions.  Namely, for our basis we use the least 
dissipative modes introduced in sec. \ref{eigmodes}. 

\subsection{Discretisation} 
As mentioned in section \ref{eigmodes}, the non-linear terms cannot be treated under 
their expanded form, or at least, the costs of doing so would be prohibitive 
\cite{canuto06_1}. Instead, they should be dealt with in the "physical" space, 
and this requires to know their value in a discrete number of points. For this, 
we start from the more general form of eq.~(\ref{invlapl}):
\begin{equation}
    \duv + \opl (\uv) = P(G(\uv)) \,,
\label{first}
\end{equation}
with appropriate boundary conditions.  $\opl$ is here  a linear operator and $G$
includes the nonlinear terms. In this section we provide a general method to 
solve this type of equations on a set of discrete points $\{\mathbf{x}_i\} = \{(x_i, y_i, z_i)\}$. The method 
does not depend on specific locations of $\{\mathbf{x}_i\}$. Therefore
we shall not specify them at this point, but only later when the method is applied to the specific problem of MHD channel flows.
We first rewrite eq. (\ref{first}) 
in the weak form:
\begin{equation} 
\langle \duv + \opl (\uv), \vv \rangle = \langle P(G(\uv)), \vv \rangle \,,
\label{weak}
\end{equation}
where $\{\vv\}$ is a set of test functions and $\langle\cdot|\cdot\rangle$ is the
scalar product associated to the $\mathcal L_2$ norm.  As we are using a
spectral approach to solve  (\ref{first}), the solution is expressed as a
series:
\begin{equation} 
\uv = \sum_n a_n \ev \label{base} \,,
\end{equation}
where $\{\ev\}$ is a functional basis, here chosen as a
set of eigenfunctions of $\opl$. If we choose as trial functions the
set of Dirac-delta functions located at points $\{x_i\}$, insert expression
(\ref{base}) into equation (\ref{weak}), and use the fact that $\ev$ are
eigenvectors of operator
$\opl$, we obtain a set of equations:
\begin{equation}
\sum_n {\dot{a}_n \mathbf{\evni} + a_n \lambda_n \mathbf{\evni}} = P(G(\uvi)) \,,
\label{LU}
\end{equation}
where $\lambda_n$ is the eigenvalue associated to $\mathbf{e_n}$. If we
express the non linear terms in the same basis, 
\begin{equation}
G(\uvi) = \sum_n g_n \ev \label{gbase} \,,
\end{equation}
we can easily calculate projection $P(G)$ by zeroing coefficients $g_n$
corresponding to irrotational modes.  Using this expansion and  the fact that
${e_i}$ are linearly independent, we can deduce from (\ref{LU}) a set of
ordinary differential equations
for the unknown coefficients $a_n$:
\begin{equation}
    \dot{a}_n = -\lambda_n a_n + g_n\,.
\label{an}
\end{equation}
The set of equations (\ref{an}) constitutes the set of discrete equations we
were looking for. Note that this method requires the number of collocation
points $\{\mathbf x_i\}$ to match the dimension of the functional basis.
\subsection{Nonlinear terms}
\label{s:nonlin}
The main difficulty in solving (\ref{an}) lies in calculating coefficients
$g_n$. 
In standard spectral methods, this is done by transforming velocity field back
to its physical expansion, calculating $\vel \otimes \vel$ there, transforming
the result of this operation forward and calculating derivatives using
properties of the basis elements. 
Unlike for Fourier modes, no simple analytical formula expresses the spatial 
derivatives of basis vectors in terms of these vectors. We therefore have to 
first calculate the physical expansion of all spatial derivatives, then 
combine them to calculate non-linear terms and 
finally calculate the spectral expansion of the result. An additional obstacle 
is that no fast transform is available to calculate $a_n$ for the basis we are 
using. Moreover, even in the relatively simple configuration considered here, 
the eigenvectors used as a basis are
not orthogonal to each other. Therefore the spectral decomposition can not be
obtained by calculating the scalar product of the decomposed field with the
elements of the basis.\footnote{It is worth mentioning, that even if the 
aforementioned basis, known analytically, was orthogonal for scalar product 
$\langle \cdot|\cdot\rangle$, we would still be facing 
the same problem with its discrete counterpart: in general it is not possible 
to find a set of discretisation points such that the associated discrete functional basis is orthogonal. 
}  Projection would require the knowledge of the eigenvectors of the adjoint operator
which cannot be readily obtained. The reverse transformation consists of the
reconstruction of a physical expansion of vector field in a given set of points
of the physical space assuming its spectral expansion is known. By
contrast to the forward transform, it is straightforward to calculate, since the values of the
basis functions are known at each of the discretisation points. 
These values can be calculated once at the beginning of calculations and used
later in an array format at every time step. This speeds up the calculations at
the cost of an increased memory usage. On this grounds, the forward
transformation (calculation of a spectral decomposition of a vector field which
is known at the set of discrete points in space) can be formulated as a set of
linear equations for unknown spectral components:
\begin{equation}
    g_n \textbf{e}_n(\textbf{x}_i) = G(\textbf{u}(\textbf{x}_i)) \,.
    \label{LUset}
\end{equation}
As the elements of matrix $\textbf{e}_n(\textbf{x}_i)$ in this set of equation do not change  during a single
numerical run, it is worth performing the $LU$ decomposition of the corresponding
matrix at the beginning of calculations and later use it to efficiently find
the spectral decompositions. 

For the above method to converge in the limit of $n \to \infty$, the
space spanned by the basis $\mathbf{e}_n$ has to include the  images of
all divergent free fields $\uv$ by the non-linear operator $G$. 
However, the eigenmodes of $\opl$ given
by \cite{dyp09_tcfd} only span the divergence free subspace of all
functions satisfying the boundary conditions of the problem. Since the 
non-linear terms do 
not preserve the divergence-free property, even when it only reduces to 
$G (\textbf{u})= \vel \cdot \nabla \vel$,  the solenoidal modes have to be complemented by 
additional non-solenoidal  
basis elements satisfying the boundary conditions. By virtue of the existence 
and uniqueness of the Helmholtz decomposition, it suffices to 
supplement the original basis with additional elements spanning the subspace 
of irrotational functions. These are obtained by solving the
eigenvalue problem (\ref{eigen}), with condition (\ref{divfree}) replaced with
$\nabla \times \vel =0$, again with the problem's boundary conditions. Their 
full form can be found in \ref{appeq}.
\subsection{Using Fourier Transform in $XY$ directions}
\label{s:fftuse}

The technique presented so far is quite general, and can be used for any
problem, as long as the solutions to the eigenvalue problem for the underlying
operator can be found and divided into two disjoint subsets: irrotational and
solenoidal. However, as seen in \ref{appeq}, the eigenmodes of $\opl$ can be
factorized as the product of two scalar functions of $x$ and $y$ respectively,
and a vector function of $z$.  Moreover, the functions of $x$ and $y$ consist
of Fourier modes, so the set of eigenmodes can be enumerated by a tuple of three numbers $(n_x,
n_y, n_z)$ and for every mode we can define the vector function $\mathbf{E}_{n_x, n_y, n_z}(z)$
such the mode takes the form
\begin{equation}
    \mathbf{E}_{n_x, n_y, n_z}(z) \exp{(i k_{n_x} x + i  k_{n_y} y)} \,.
\end{equation}
We can thus speed the transformation from physical to spectral space by first 
performing two dimensional Fast Fourier Transform in the $x-y$ direction. This brings the transformed field
under the form:
\begin{equation}
    G(\vel(x_i, y_i, z_i) = \sum\limits_{n_x, n_y} \mathbf{A}_{n_x, n_y}(z_i) \exp{(i 2\pi n_x x_i + i 2\pi n_y y_i)} \,,
\end{equation}
where $\mathbf{A}_{n_x, n_y}$ is the complex amplitude of Fourier mode $(2\pi n_x, 2\pi n_y)$.
Then, following the technique described in \ref{s:nonlin}, for every value of $(n_x, n_y)$
we find the set of spectral coefficients $\{g_{n_x, n_y, n_z}\}$ by solving a set of equations
\begin{equation}
    g_{n_x, n_y, n_z} \textbf{E}_{n_x n_y n_z}(z_i) = \mathbf{A}_{n_x, n_y}(z_i) \,.
\end{equation}
The reverse transformation is speeded up in a similar
fashion. 

Using the Fast Fourier transform in $x-y$ planes imposes the distribution of
discretization points in these planes: they have to form a regular rectangular
grid. We denote its dimensions as $N_x\times N_y$. In our simulations we also use a uniform grid in $z$ direction of dimension $N_z$. In order for set of equations (\ref{LUset}) to have a unique solution the number of modes used during the spectral decomposition has to be equal $N_z$, while the total
number of independent modes used in the caluclations is $N=N_x N_y N_z$.
\subsection{Estimation of number of required discretization points in $z$ directions}
\label{s:nmodes}
\begin{figure} \setlength{\unitlength}{1cm} \begin{center}
    \includegraphics[width=11cm]{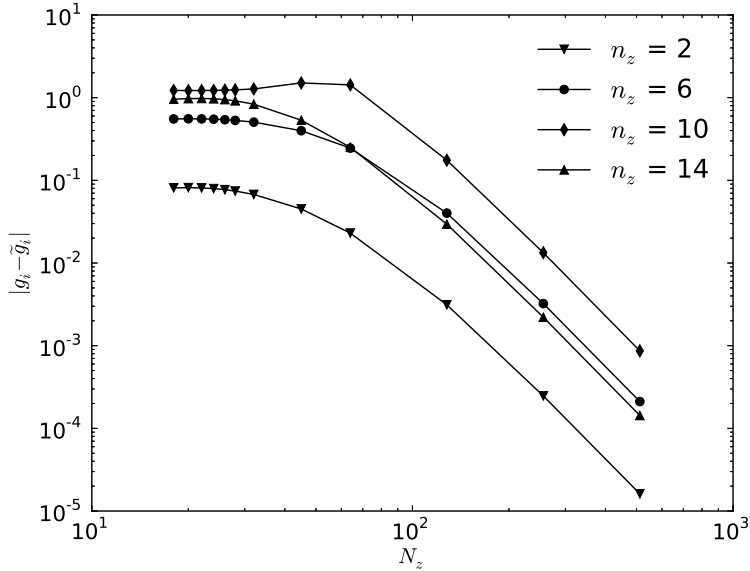} \end{center}
\caption{The relative error in determination of $g_{8,0,n_z}$
for $n_z=2, 6, 10, 14$ as a function of grid size in the $z$ direction. The
decomposed field is $(\vel \cdot \nabla) \vel$, where $\vel$ is the single
eigenmode $\textbf{e}_8$ for $\ha=1$ normalised in such a way that its amplitude is 1.}
\label{error_small} \end{figure}

\begin{figure} \setlength{\unitlength}{1cm} \begin{center}
    \includegraphics[width=11cm]{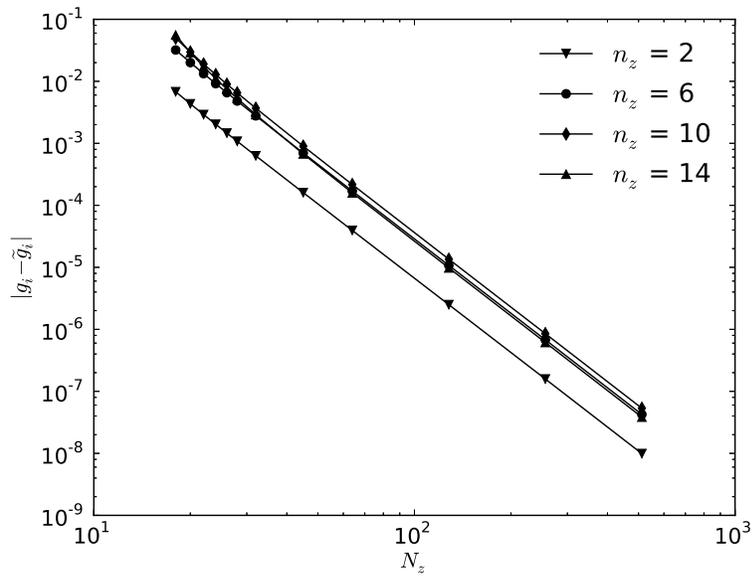} \end{center}
\caption{The relative error in determination of $g_{8,0,n_z}$
for $n_z=2, 6, 10, 14$ as a function of grid size in the $z$ direction. The
decomposed field is $(\vel \cdot \nabla) \vel$, where $\vel$ is the single
eigenmode $\textbf{e}_8$ for $\ha=100$ normalised in such a way that its amplitude is 1.}
\label{error} \end{figure}

The technique described above has the advantage that the obtained spectral
decomposition reproduces exactly the physical field on the given set of
discretisation points. Therefore momentum and energy are conserved by this
procedure. However the obtained spectral coefficients $g_n$ are different from
the exact ones $\widetilde{g}_n$, which would be the result of decomposition of
the same vector field in a space of infinite dimension spanned by all
eigenvectors.$|\widetilde{g}_n-g_n|$ is the so-called aliasing error. In spectral methods
based on standard functional bases (Fourier or orthogonal polynomials),
there are known procedures to correct this error.  They all rely on the fact that the
spectral decomposition of non-linear terms can be calculated analytically for
elements of the functional basis. Unfortunately in our case these are only known  in the $x$ and $y$ directions,
where we adopt the $3/2 N$ rule for dealiasing \citep{canuto06_1}. In the $z$ 
direction we adapt this procedure by performing the discrete transformation 
with additional number of modes $N$ larger than the one strictly required by 
the system's dynamics, $N_D$ (The latter is of the order of the attractor 
dimension of the dynamical system underlying the given problem \cite{pa06_pf}). 
After every evaluation of the spectral decomposition, the coefficients 
corresponding to these  additional modes are padded with zeros. In other words, 
the set of equations (\ref{LUset}) is expanded over $N$ modes, but only the 
first $N_D$ coefficients are calculated. Since the decomposition over $N$ 
modes is more precise than that over $N_D$ modes but still not exact, this 
procedure does reduce the dealiasing error but does not remove it completely. 
Note that the number of discretisation points is then $N$, not $N_D$.\\
We have evaluated the residual aliasing error on a number of examples  and
reported some of them in fig.(\ref{error}) and (\ref{error_small}). These show
the relative error $\epsilon_a=|g_n - \widetilde g_n|/g_n$ for several values
of $n$ as a function of the number of modes used in decomposition in the $z$
direction $N_z$. The decomposed field is a term corresponding to non linear
terms $(\vel \cdot \nabla) \vel$, where $\vel = \textbf{e}_{8}$, which we found
to be representative of the behaviour observed using other modes. \\
Two regimes can be distinguished. For low values of $\ha$, the relative error in 
$g_n$ drops monotonically for all plotted values of $N_z$ (see fig. 
\ref{error_small}). This is because the smallest scales in the decomposed field 
result from its oscillatory part, which physically represent small vortices in 
the bulk of the flow, the size of which is practically independent of $\ha$ 
(see \cite{dyp09_tcfd}). 
The picture looks qualitatively different at high Hartmann numbers 
(see fig. \ref{error}). In this case, the smallest structures in the decomposed 
velocity field are the Hartmann layers. Up to $N_z \sim \ha$, the number of 
discretisation points is insufficient to resolve them. Consequently, the
dealiasing errors is high and remains more or less constant. Above this value
of $N_z$, the Hartmann layer is correctly resolved and the error introduced by 
spectral decomposition starts to drop, roughly following a power law of the 
form $\epsilon_a\sim N_z^{-\alpha}$, with $\alpha \approx 4$. We use this
approach in every calculation to determine the number of discretisation points
required in the $z$ direction in order to keep the 
dealiasing error below the specified level of accuracy $\epsilon$: for a given
value of $N_D$, this is done by constructing the non-linear term
\[
\textbf{e}_{N_D} \cdot \nabla \textbf{e}_{N_D} \,,
\]
calculating its spectral decomposition for different values of $N$, and pickup the smallest value $N$ for which the relative error $\epsilon_a$ is below 
$\epsilon/N$. Since modes of highest index are expected to generate the largest 
error, this ensures that dealiasing remains below the specified precision.
%
%
\subsection{Parallelisation}
 The code has been parallelised for memory distribution systems using the domain
decomposition approach. Since in order to calculate Fourier transform in the
$x$ and $y$ direction we are using the \textsc{FFTW} library \cite{fftw}, we
adopted the strategy implemented therein to distribute arrays between different computational nodes. Its principle is that every three dimensional array 
representing a component of velocity as a function of $\textbf{x}$ (in physical 
representation) or their spectral decomposition (in spectral representation) is 
divided into slabs orthogonal to their first dimension. In our problem, this 
corresponds to a division of computational domain into slabs along planes 
$x=const$ and $k_x=const$ in physical and spectral representations respectively.
Every computational core is assigned such a single block. Additional speedup is 
gained from the ability of the library to perform multiple transforms at once, 
of interleaved data. This makes it possible to decrease the cost of inter-node 
communication by calculating multiple \textsc{2D FFT} transformations for 
different values of $z$ simultaneously. Further speed up can be obtained by parallelizing 
of the matrix by vector multiplications during the transformation from physical to spectral expansion.

\subsection{Time discretization}
\label{stdis}
To numerically solve equation set (\ref{an}) of ODEs in time, we use an 
integrating factor technique (see \cite{canuto06_1}),
and rewrite eqs. (\ref{an}) as:
\begin{equation}
    \frac{d}{d t} [e^{\lambda_n t} a_n] = - e^{\lambda_n t} g_n \,.
    \label{tdisc}
\end{equation}
We then implement the Euler approximation to provide solution $a^{k+1}_n$ at 
instant $t_{k+1}$ in terms of solutions $\textbf{a}^k_n$ at previous timestep 
$t_k = t_{k+1}-\Delta t$
\begin{equation}
    a^{k+1}_n = - e^{\lambda_n \Delta t} [a^k_n - \Delta t\,g_n(\textbf{a}^n)] \,.
\end{equation}
The value of $\Delta t$ was chosen at every time step according to the equation:
\[
\Delta t = 1.2 \min\limits_n \left|\frac{a_n}{g_n(\textbf{a}^n)}\right| \,.
\]
 
\section{Comparison with DNS}
\label{sec:tests}
\subsection{Reference numerical setup based on \opf}
To validate the present numerical scheme we compared the results it produces
with those of three-dimensional, time-dependent direct numerical simulations
performed with a code based on the open source framework \opf, on
test cases of freely decaying MHD turbulence.  \opf is based on
the finite volume approach and uses a co-located grid. Our MHD code is based on
it, and implements the set of low $R_m$ MHD equations
(\ref{eq:nav}-\ref{eq:ohm}) using second order spatial discretization and
second order, implicit time scheme. The equations are solved segregated and the
PISO algorithm is used to handle the velocity-pressure coupling. The
conservation of charge is ensured by calculating current density and Lorentz
force following the approach proposed by \cite{ni07_jcp}.  The details of its
implementation in \opf and its validation are described by
\cite{dp12_jfm}. The spectral methods described in the previous sections were
implemented by modifying the spectral code TARANG developed by
\cite{verma13_pramana}.\\
Our numerical domain was a cube of dimension $2L$ divided uniformly into $N$
cells in every direction. This configuration is a fair representation of the
experiment conducted in \cite{kp10_prl}, in which MHD turbulence is
electrically driven in a cubic container pervaded by a transverse magnetic
field. In order to calculate correctly the electric current density in Hartman
layers we always resolve each of them with at least three computational cells
in the $z$ direction as in \cite{dp12_jfm}. Therefore the Hartman number is
connected to the resolution by $N=\ha/6$. Following the DNS of decaying MHD
turbulence in a three-dimensional periodic domain by \cite{okamoto10_jfm}, the
initial conditions consist of a random gaussian velocity field with $u(k) \sim
\exp{[(-k/k_p)^2]}$ where $k_p = 4 \pi / L$.  This corresponds to the energy
spectrum $E \sim k^4 \exp{[-2(k/k_p)^2]}$.  For this choice of initial velocity
field, the integral scales of turbulent motions is given by $l=\sqrt{2
\pi}/k_p$.  The velocity spectrum was normalised in such a way that cell sizes
correspond to $l_K/1.4$ where $l_K = l Re^{-3/4}$ is the Kolmogorov length
scale and the Reynolds number in its definition $Re=u^\prime l/\nu$ is based on
$l$ and velocity $u^\prime=u(k=k_p)$. With this choice, the Reynolds number and
the Hartmann number are linked by $Re= \mathrm{0.33} \ha^{4/3}$. This strategy
allows us to calculate the most intense flow possible whilst minimizing
mesh-induced numerical errors at a given mesh size, since the mesh is always
uniform. 
\subsection{Simulations at low $\ha$}
For the reference case, we have chosen $\ha=56$, a value within reach with 
a traditional code such as \opf. Adopting the
procedure presented in previous subsection for \opf calculations we use
 $N_x = N_y = N_z =170$ number of points in every direction and initial
conditions characterized by Reynolds number $Re=28$. We have followed the
evolution of the initial conditions up to the time corresponding to $30\,t_J$
where $t_J=\sigma B^2/\rho$ is the timescale of Joule dissipation. We have also
performed a shorter calculation up to $t=t_J$ with twice this resolution in every direction. 
The difference between the magnetic and viscous dissipation rates were smaller 
then 1 per cent, so it is safe to consider that the run in standard resolution 
is fully converged.

For the corresponding spectral calculations we use a resolution 1.5 times
higher ($N_x = N_y = N_z =256$) in order to reduce the dealiasing errors. The
initial conditions were chosen in such a way that their physical expansion on
grid $N_x =N_y = N_z =170$ was identical to the initial conditions used in the
calculations with \opf.

\begin{figure}
    \label{fig:diss}
    \includegraphics[angle=-90,width=11cm]{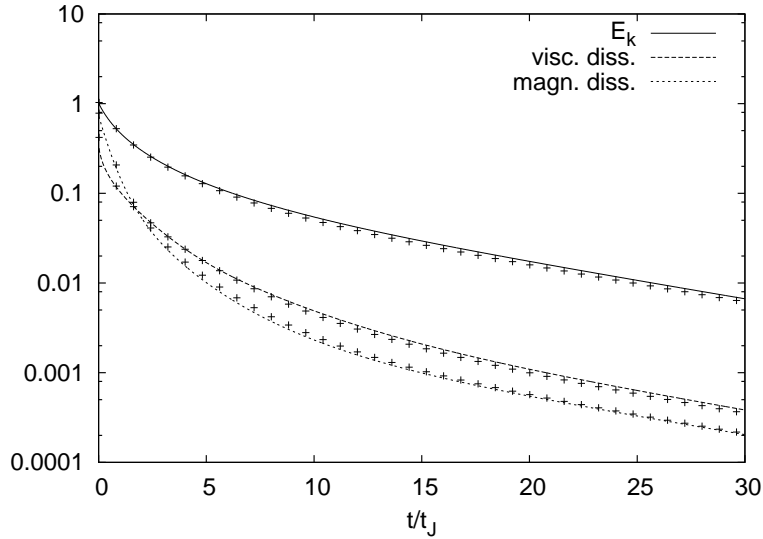}
    \caption{Total kinetic energy (normalized by its initial value) and viscous and magnetic dissipation rates (normalized by initial
    kinetic energy divided by $t_J$) in test case with $\ha=56$ in the function
    of Joule times. The lines and points 
    respectively represents the results obtained with finite volume and spectral codes.}
\end{figure}
The fig. \ref{fig:diss} shows the evolution of normalized global kinetic
energy, the viscous and magnetic dissipation rates \emph{vs.} time normalised
by the Joule time. This quantities were respectively calculated as:
\begin{eqnarray}
    E(t)=\int \frac{1}{2} \vel^2 \mathrm{d}V\,, &
    \epsilon_\nu= \int \frac{1}{2} \sum\limits_{ij} \sigma_{ij} \sigma_{ij} \mathrm{d}V\,,&
    \epsilon_M=\int \sigma \mathbf{j}^2 \mathrm{d}V\,,
\end{eqnarray}
where
\begin{equation}
    \sigma_{ij} = \nu (\frac{\partial u_i}{\partial x_j} + \frac{\partial u_j}{\partial x_i})
\end{equation}
is the viscous stress tensor.  The normalization factor corresponds to the
initial value of kinetic energy for $E$ and to the initial value of of kinetic energy multiplied by Joule time for dissipation rates.. 
All presented quantities exhibit quantitatively and qualitatively the same
agreement in both codes.  The spectral code exhibits a slightly smaller values
of viscous dissipation then the finite volume code and slightly higher magnetic
dissipation.  The relative difference in the total kinetic energy after 30
Joule times is below $10\%$, the spectral code being overall the more
dissipative one.  

\begin{figure}
    \label{fig:lambda}
    \includegraphics[angle=-90,width=11cm]{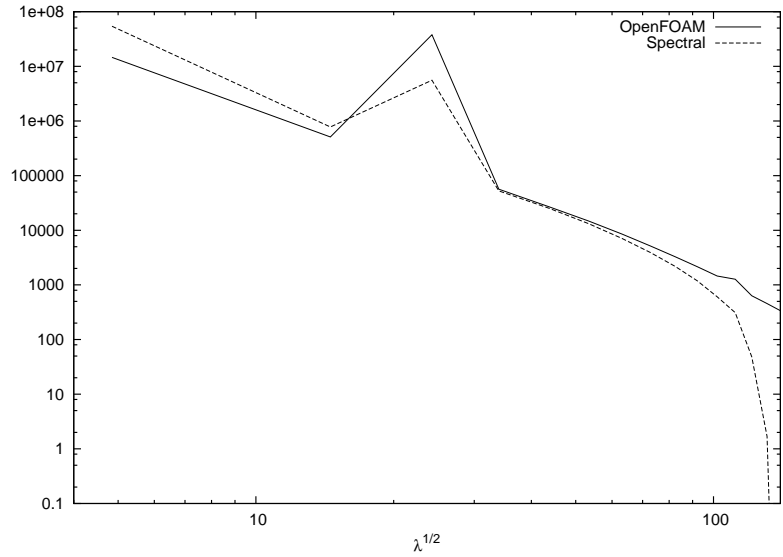}
    \caption{The energy contained in modes with given value of $\lambda$ for case with $\ha=56$ after $t=30 t_J$. The solid line represents
    calculations performed with \opf while dashed line shows the results obtained with the spectral code.}
\end{figure}

We also compared the power spectral densities for spectral parameter
$\sqrt{-\lambda}$ resulting from both codes at $t=30\tau_J$ ($\sqrt{-\lambda}$
is the equivalent of the wavenumber in Fourier based spectral decomposition \cite{dyp09_tcfd}).
To this aim we expanded both numerical solutions onto the set of least 
dissipative modes used in the spectral simulation, following the procedure 
described in section \ref{sec:method} and compared the kinetic energies
corresponding to each value of $\lambda$. The results 
are presented in fig. \ref{fig:lambda}. Both codes produce quantitatively 
similar results.
Qualitatively, the spectral code seems to be more dissipative, and as a result,
more energy is concentrated in the largest scales at the expense of the
large-intermediate ones. For mid to small scales up to $|\lambda| \sim 80$ both codes agree well. The
difference in the smallest scales can be explained as follows. The physical space of
collocation points is rectangular, so is the spectral domain in the $(k_x, k_y,
\kappa_z)$ space. Sine curves $\lambda(k_x, k_y, \kappa_z) = const$ are a family of cardioids, this space only includes full $\lambda$ - shells up to
$|\lambda| \approx 80$. On these grounds the $\sqrt{\lambda}$ - PSD spectra is a
physically meaningful representation of the solution up to $|\lambda| = 80$.
\begin{figure}
    \label{fig:big}
    \includegraphics[angle=-90,width=11cm]{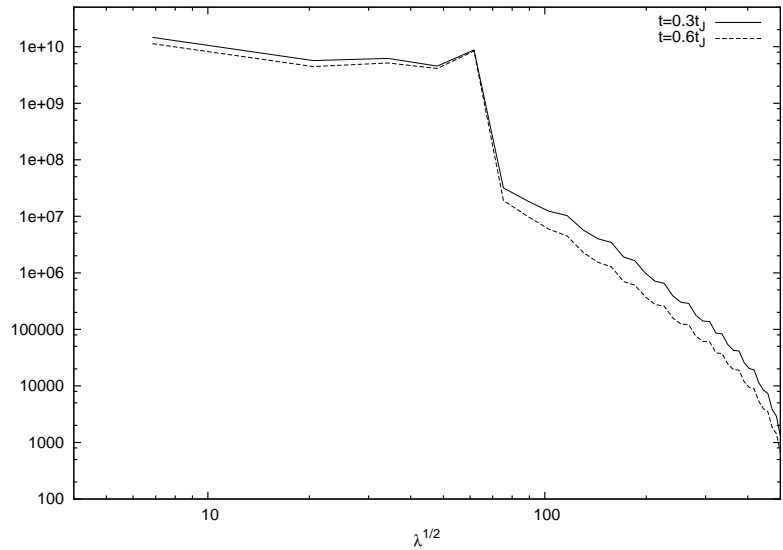}
    \caption{The energy contained in modes with given value of $\lambda$ for case with $\ha=56$. The solid line shows $\lambda$-spectra at $t=0.3$ Joule times. Then all the modes with $\lambda>100$ were padded to zero and the flow was evolved for another $0.3$ Joul times.
    The final $\lambda$ spectrum is represented by dashed line.}
\end{figure}
\subsection{Simulations at high $\ha$}
In the previous section we showed that the results obtained with the new 
spectral approach were able to reproduce the same as those obtained with 
 using finite volume technique, in a case accessible to both methods. For the 
purpose of this test, we used in our 
spectral calculations and an effective number of modes $N_D$ 
(those which were not padded with zeros in the dealiasing procedure), which was the same as the number of  grid cells in calculations performed with 
\opf. This imposed 
to choose a rather low Hartmann number so as to keep the computational cost low 
is both methods. It is, however, for flow at high $\ha$, that  our new
approach can potentially offer computational savings, because the 
theoretically predicted number of degree of freedom is much lower than the
number of grid points required in the mesh used in the finite volume method (or
in a classical spectral method).  The full calculations of decaying or forced
MHD turbulence at high $\ha$ are beyond scope of this contribution, as they would
require a full scale parallelisation of our code, which is not yet implemented.
Our main purpose is to present an implementation of our new spectral method and
show that it gives accurate results. Nevertheless, we shall still demonstrate
its potential for calculations at higher values of $\ha$.  To this end, we used
a different approach to examine the computational savings offered by this new
technique. 
We start with initial conditions determined as in the previous section for 
$\ha=108$. The flow is then evolved for a period of $0.3 t_J$ using a 
spectral basis with a number of modes sufficently high to resolve \ldots so as to erase the trace of the 
initial non-MHD flow. At this instant we 
removed from the resulting velocity field all contributions from modes with
$|\lambda| > |\lambda_{\rm max}| = 10^4$. The value of $\lambda_{\rm max}$ was
chosen as \ldots the value of the value predicted by
\cite{pdy10_jfm} for the smallest scales in the forced flow with the forcing on
scales $2\pi/k_p$ and Reynolds number $Re(0.3\,t_J)=$ This value is located
just behind a sharp drop in PSD, which is an early indication of its relevance
as an estimate for the smallest scales.
Subsequently we let the flow evolve
for another $0.3\,t_J$. We then we examined the Power density
$\lambda$-spectrum of the resulting velocity field and compared it to that at
$t=0.3 t_J$, 
seeking in particular if energy had leaked into regions of the spectrum where 
$|\lambda| > |\lambda_{\rm max}|$. The results are presented on fig. \ref{fig:big}
We see that the evolution of the spectrum was not significantly influenced.
This indicates that the evolution of the flow is sufficiently well described by
the modes with $|\lambda| \leq |\lambda_{\rm max}|$, and our method does carry the ability to
resolve high $\ha$ flows with the limited number of modes.
\section{Computational cost}
The cost of pseudospectral codes based on FFT technique scales as  $N_x N_y N_z \log(N_x
N_y N_z)$.  In DNS calculations $N_i$ should be of the order of Reynolds number $Re^{3/4}$
each. Moreover, to obtain physically meaningful results it is necessary to
resolve the Hartmann layer in $z$ direction. $N_z$ should therefore be at least
of the order of $\ha$. For large Harmann number calculations the
computational cost of FFT based code thus scales as $\sim Re^{3/2} \ha \log(Re \ha)$.
As explained in sect. \ref{s:fftuse}, the
spectral decomposition described in sect. \ref{s:nonlin} is only used to treat $z$-dependence
while the spectral decomposition in $x-y$ can be
performed with FFT.  The computational cost in this case scales as $\sim N_x
N_y N_z^2$. Using the estimation for the number of degrees of freedom in low Rm MHD flows from
\cite{pa03,pa06_pf}: $N_{x,y} \sim Re^{1/2}$, $N_z \sim Re/\ha$. In sect
\ref{s:nmodes}, however, we found that it is still necessary to have $N_z \sim
\ha$ to calculate accurate spectral coefficients with small errors. Nevertheless,
we only need to calculate $Re^2/\ha$ spectral coefficients, as the rest of
them will be dissipated
on the very short timescales and are not physically meaningful. Therefore the
total computational cost of the proposed scheme scales as $\sim Re^2$. The proposed method
has the potential to significantly reduce the computational cost incurred by traditional ones in the regime of high values of $\ha$, as
its computational cost does not depend on the intensity of the magnetic field,
which a limiting factor for traditional methods. Theoretically, there is room
for additional improvement as the number of degrees of freedom in low Rm MHD
flows is inversely proportional to $\ha$.  However, this would require an
alternative approach for the treatment of the non-linear terms.
\section{Conclusions}
We have presented a novel model method to simulate MHD flows in
channel configuration. It is based on using the sequence of least dissipative
eigenmodes from the dissipation operator instead of the traditional Fourier or
Tchebychev basis.  The advantage is that these modes show physical properties
of the actual MHD flow. Consequently, the basis requires a smaller number of
modes. On the other hand, they lack the advantageous properties of usual bases, 
such as orthogonality and the existence of fast transforms, 
which are heavily relied upon in standard spectral methods. We have put forward an approach to 
overcome this difficulties and were able to construct a useful numerical 
scheme based on this alternative modes. We tested this new method by comparing
its ability to calculate the freely decaying MHD turbulence in a channel to
that of a finite volume code and obtained very good agreement between the two
codes both in terms of global quantities like energy dissipations rates and in
terms of the resulting energy spectra. As a last test, we verified the potential of the
new scheme to efficiently resolve flows in the presence of high magnetic fields
with a limited number of modes.\\

We would like to stress the large potential of the method we presented.
The scheme we put forward doesn't depend on any special properties of the used
modes (in particular they don't have to be orthogonal).  Therefore the
procedure can be easily adapted to other MHD and non-MHD problems. The main
idea is to use a basis of functions that already exhibit those features of
the flow, that incur heavy computational cost beyond that
strictly required by the underlying dynamical system (boundary layers typically).
This makes it possible to save the computational time of having to reconstruct
these structures with elements of more standard basis.  In favourable cases,
where this operation is computationally expensive (typically when boundary
layers are very thin), this technique has a potential to speed up DNS
calculations by an important factor by restricting the spectral domain to one
more closely embedding that relevant to the calculated flow.\\

The authors gratefully aknowledge financial support of the Leverhulme Trust
(Grant F00/732J). The authors are also grateful to Professor Mahendra Verma,
for making the TARANG code available to them and for fruitful discussions
around the ideas presented in this paper. Finally the authors would like to express their gratitude to
Antoine Huber for his work in the numerical simulations carried out with \opf.

\appendix
\section{Analytical expression of the eigenmodes of $\opl$}
\label{appeq}
\cite{dyp09_tcfd} provides solutions to divergence free eigenvalue problem for operator (\ref{eigen}). The eigenvalues
$\lambda$ are given by dispersion relation:
\begin{equation}
\lambda=-(k_x^2+k_y^2-K)+\ha^2\frac{K}{k_x^2+k_y^2-K}.
\label{eq:disp}
\end{equation}
where for a given $\lambda$ $K$ can take values $-\kappa_z^2$ and $\mu^2$. Therefore they are related by:
\begin{equation}
{\kappa_z^2}{\mu^2}-k_\perp^2(k_\perp^2-\mu^2+\kappa_z^2 
+\ha^2) =0.
\label{eq:kzd}
\end{equation}

\subsection{Divergence-free modes}
Divergence free eigenvectors can be divided into three subgroups.

\subsubsection{Orr-Sommerfeld modes}
These are in turn divided based on whether velocity component perpendicular to $\mathbf{e}_z$ is symmetric or antisymmetric in $z$.

Symmetric modes:
\begin{equation}
\label{eq:orrmode1}
\begin{aligned}
\mathbf e^{OSs}_\lambda&=
\bigg\{
i\kappa_z \cos{\kappa_z} \mathbf k_\perp
\left(\frac{\cos(\kappa_zz)}{\cos(\kappa_z)}-\frac{\cosh(\mu z}{\cosh(\mu)}\right) \\ 
&+k_\perp^2 \sin{\kappa_z} \mathbf e_z 
\left(\frac{\sin(\kappa_zz)}{\sin(\kappa_z)}-\frac{\sinh(\mu z)}{\sinh(\mu)}\right)
\bigg\} \exp(i\mathbf k_\perp\cdot\mathbf r_\perp).
\end{aligned}
\end{equation}
with $\kappa_z$ and $\mu$ satisfying:
\begin{equation}
\label{eq:orr_s}
\kappa_z\tanh(\mu)-\mu\tan(\kappa_z)=0
\end{equation}

Antisymmetric modes
\begin{equation}
\label{eq:orrmode2}
\begin{aligned}
\mathbf e^{OSa}_\lambda&=\bigg\{
i\kappa_z \sin{\kappa_z} \mathbf k_\perp
\left(\frac{\sin(\kappa_zz)}{\sin(\kappa_z)}-\frac{\sinh(\mu z)}{\sinh(\mu)}\right)\\
&-k_\perp^2 \cos{\kappa_z} \mathbf e_z
\left(\frac{\cos(\kappa_zz)}{\cos(\kappa_z)}-\frac{\cosh(\mu z)}{\cosh(\mu)}\right)
\bigg\}
\exp(i\mathbf k_\perp\cdot\mathbf r_\perp).
\end{aligned}
\end{equation}
with $\kappa_z$ and $\mu$ satisfying:
\begin{equation}
\label{eq:orr_type2}
\kappa_z\tan(\kappa_z)+\mu\tanh(\mu)=0
\end{equation}

\subsubsection{Square modes}
Symmetric modes:
\begin{equation}
\label{eq:squiremode_s}
\mathbf e^{Ss}_\lambda= \mathbf k_\perp\times\mathbf e_z
\left(\frac{\cos(\kappa_zz)}{\cos(\kappa_z)}-\frac{\cosh(\mu z)}{\cosh(\mu)}\right)
\exp(i\mathbf k_\perp\cdot\mathbf r_\perp).
\end{equation}
with $\kappa_z$ and $\mu$ satisfying:
\begin{equation}
\label{eq:squire_s}
S\tan(\kappa_z)+M\tanh(\mu)=0
\end{equation}

Antisymmetric modes
\begin{equation}
\label{eq:squiremode_as}
\mathbf e^{Sa}_\lambda= \mathbf k_\perp\times\mathbf e_z
\left(\frac{\sin(\kappa_zz)}{\sin(\kappa_z)}-\frac{\sinh(\mu z)}{\sinh(\mu)}\right)
\exp(i\mathbf k_\perp\cdot\mathbf r_\perp).
\end{equation}
with $\kappa_z$ and $\mu$ satisfying:
\begin{equation}
\label{eq:squire_as}
S\tanh(\mu)-M\tan(\kappa_z)=0
\end{equation}

Quantities $S$ and $M$ are defined by relations:
\begin{displaymath}
S=\kappa_z \left(k_\perp^2 - \mu^2\right) \quad {\rm and} \quad
M=\mu \left(k_\perp^2 + \kappa_z^2\right) .
\end{displaymath}

\subsubsection{Modes with $\mathbf k_\perp$=0}
These are the modes in which velocity is function of $z$ coordinate only . They are
omitted in \cite{dyp09_tcfd} and we had to derive them independently. Because of the requirement that $\vel$ is divergence free
and the non-slip boundary conditions at $z=\pm1$, $u_z$ has to vanish everywhere. Moreover $\nabla \cdot (\sigma \vel \times
\mathbf{B})=0$ for this modes. Additionally $\sigma \vel \times \mathbf{B}$ satisfies boundary conditions imposed on $\mathbf{j}$.
Therefore in this case the electric potential vanishes in the whole of the domain for the considered modes and the electric current 
density is given by the expression:
\[
\mathbf{j} = \sigma \vel \times \mathbf{B}
\]
Therefore the eigenvalue problem in this case reduces in dimensionless form to:
\begin{eqnarray*}
 \Delta \vel &=& (\lambda + \ha^2) \vel\\
 \left. \vel \right|_{z=\pm 1} &=& 0
\end{eqnarray*}
Again the modes can be divided into symmetric and antisymmetric ones.

Symmetric modes:
\begin{equation}
    \mathbf{e}^i_{\kappa_z} = \cos (\kappa_z z) \mathbf{e}_i \qquad i=x,y
\end{equation}
where $\kappa_z = \pi/2\,q$ and q is an odd natural number

Antisymmetric modes:
\begin{equation}
    \mathbf{e}^i_{\kappa_z} = \sin (\kappa_z z) \mathbf{e}_i \qquad i=x,y
\end{equation}
where $\kappa_z = \pi/2\,q$ and q is an even natural number

\subsection{Irrotational modes}
Our approach required also derivation of curl-free eigenvectors. They are in one to one relation with Orr-Sommerfeld modes.

Symmetric modes:
\begin{equation}
\label{eq:sols}
\begin{aligned}
\mathbf e^{Cs}(\lambda)&=
\bigg\{
\cos{\kappa_z} \mathbf k_\perp
\left(\frac{\cos(\kappa_zz)}{\cos(\kappa_z)}-\frac{\cosh(\mu)}{\cosh(\mu)}\right)\\
&+i\kappa_z  \sin{\kappa_z} \mathbf e_z 
\left(\frac{\sin(\kappa_zz)}{\sin(\kappa_z)}-\frac{\sinh(\mu z)}{\sinh(\mu)}\right)
\bigg\}
\exp(i\mathbf k_\perp\cdot\mathbf r_\perp).
\end{aligned}
\end{equation}
with $\kappa_z$ and $\mu$ satisfying (\ref{eq:orr_type2})

Antisymmetric modes:
\begin{equation}
\label{eq:sola}
\begin{aligned}
\mathbf e^{Ca}(\lambda)&=
\bigg\{
-\sin{\kappa_z} \mathbf k_\perp
\left(\frac{\sin(\kappa_zz)}{\sin(\kappa_z)}-\frac{\sinh(\mu z)}{\sinh(\mu)}\right)\\
&+i\kappa_z \cos{\kappa_z} \mathbf e_z 
\left(\frac{\cos(\kappa_zz)}{\cos(\kappa_z)}-\frac{\cosh(\mu z)}{\cosh(\mu)}\right) 
\bigg\}
\exp(i\mathbf k_\perp\cdot\mathbf r_\perp).
\end{aligned}
\end{equation}
with $\kappa_z$ and $\mu$ satisfying (\ref{eq:orr_s})

Solutions to dispersion relations are such that for a given $\mathbf{k_perp}$ and every $q \in \mathbb{N}$, there are six
modes for which $\kappa_z \in [q\pi, (q+1)\pi]$.

\subsection{Distribution of the modes}
For a given $\mathbf{k_\perp}$ and every $q\in \mathbb{N}$ there are six modes with $\kappa_z \in [q \pi, (q+1)\pi]$,

\bibliography{biblio}
\bibliographystyle{elsarticle-num-names}
\end{document}